\documentclass[showkeys,showpacs,prl,nofootinbib,superscriptaddress,floatfix, preprint]{revtex4}
\usepackage{graphicx, amsmath, dcolumn, setspace, latexsym, amsfonts, amssymb, natbib}

\begin{document}

\title{Self-diffusion in binary blends of cyclic and linear polymers}
\date{\today}
\author{Sachin \surname{Shanbhag}}
\email{sachins@scs.fsu.edu}
\affiliation{School of Computational Science, Florida State University, Tallahassee, FL, 32306}
\affiliation{Department of Chemical and Biomedical Engineering, FAMU-FSU College of Engineering, Tallahassee, FL 32310}
\pacs{02.70.Uu (Applications of Monte Carlo Methods), 83.80.Tc (Polymer blends (rheology))}
\keywords{rings, cyclic, entanglement, blend, diffusion, polymer melt}

\begin{abstract}
A lattice model is used to estimate the self-diffusivity of entangled cyclic and linear polymers in blends of varying compositions. To interpret simulation results, we suggest a minimal constraint release model for the motion of a cyclic polymer infiltrated by neighboring linear chains. Both, the simulation, and recently reported experimental data on entangled DNA solutions support the simple model over a wide range of blend compositions, concentrations, and molecular weights.
\end{abstract}

\maketitle

\section{Introduction}

According to classical theories of polymer physics, flexible chains in solution assume coiled conformations. As the polymer concentration, or contour length is increased, these coils overlap and produce entanglement effects, which include a pronounced retardation of molecular mobility.  For linear polymers (LP), this transition from unentangled to entangled dynamics is marked by a change in the zero-shear viscosity $\eta_0$ from $\eta_{0} \sim M$ to $\eta_{0} \sim M^{3.4}$, and self-diffusivity $D_L$ from $D_L \sim M^{-1}$ to $D_L \sim M^{-2.4}$, where $M$ is the molecular weight. Molecular topology has a strong effect on the dynamics of polymers in the entangled state, and constitutes a subject of fundamental and industrial interest in polymer physics and rheology. Although important gaps in understanding persist, the behavior of entangled linear and branched polymers, such as stars, are relatively well-described  at small deformation rates, using the tube ansatz. Here, the molecular dynamics of a tracer chain enmeshed in a matrix of other molecules, are formulated in terms of a diffusion problem that describes the motion of \textit{chain ends} in a hypothetical tube \cite{doipd,degennessc,ball89}.

Concentrated solutions of ring or cyclic polymers (CP), which lack chain ends, are scientifically intriguing, since they defy a simple description in terms of the tube model. While interest in CPs has recently been rekindled, the conformational and dynamic properties of CPs in gels and in melts have been studied theoretically \cite{cates86, rubinstein86, obukhov94, iyer06}, computationally \cite{muller96, muller00a, brown01, geyler88, hur06}, and experimentally \cite{roovers85a, roovers88a, mckenna87, mckenna89, mills87, tead92, cyclicpolymers} since the 1980s. Almost all the theoretical and computational studies have investigated the characteristics of pure CPs and LPs, while blends of CPs and LPs have escaped  the same level of scrutiny \cite{geyler88, iyer07, subramaniam08}. These blend systems are important  for two reasons: (i) most experimental data on pure CPs are, in fact, data on cyclic-linear blends (CLB), due to contamination or limitations of purification methods, and (ii) the dynamics of such CLBs, are extraordinarily sensitive to the concentration of LPs, as demonstrated by the linear viscoelastic response of polystyrene CLBs \cite{mckenna86, vlassopoulos06}, and self-diffusion studies of DNA solutions \cite{robertson07a, robertson07b}. These studies indicate a dramatic change in the mobilty of CPs that is, both, unexpected and unexplained, and might supply deep insights into entangled polymer dynamics.

In this letter,  a lattice-based dynamic Monte Carlo method called the bond-fluctuation model (BFM) is used to monitor the trajectory of LPs and CPs in entangled CLBs. A minimal model is constructed, to interpret the diffusivities obtained from the simulation. This approximate theory is able to reasonably explain recent experimental data on entangled DNA solutions.

\section{Model and Methods}

We use Shaffer's version of the BFM, \cite{shaffer94, shanbhag05} which has recently been applied to blends of CPs and LPs \cite{iyer07, subramaniam08}. Since the model has been described in detail previously, only a brief summary presented here. Monomers or beads are placed on a simple cubic lattice. To generate an equilibrated CLB, we insert $n_{C}$ non-concatenated CPs and $n_{L}$ LPs, each consisting of $N$ monomers, on a 3D cubic lattice in a simulation box of size $L \times L \times L$. To simulate melt-like behavior, the total fractional occupancy of the lattice is maintained at $\phi = \phi_{C} + \phi_{L} =0.5$, where $\phi_{i} = n_{i} N/L^{3}$  represents the fractional occupancy of CPs ($i=C$) and LPs ($i=L$). A trial move is attempted by displacing a randomly selected bead, belonging to either a CP or LP, by one lattice unit. It is accepted if it does not violate the excluded volume, chain connectivity and chain uncrossability constraints. One Monte Carlo step (MCS) consists of ($n_{L} + n_{C}) N$ trial moves. Throughout this letter, length is expressed in units of lattice spacing and time in MCS.

In this study, we considered two series of CLBs, viz. $N = 150$, and $N = 300$. We varied the composition of the blend from $\phi_{C}=0.5$ to $\phi_{C}=0.0$, corresponding to the range between pure CPs to pure LPs, respectively, as summarized in Table \ref{tab_systems}. The duration of the simulation, $\tau_{sim}$ was picked to ensure that molecules had diffused at least two radii of gyration. During this period, we tracked the positions of the molecules by taking snapshots at periodic intervals. To determine the self-diffusivity, we calculated the mean-squared displacement of the center-of-mass of LPs and CPs separately, via, $g_{3}(t) = \langle  \left( \textbf{r}_{cm}(t+\tau)-\textbf{r}_{cm}(\tau) \right)^{2} \rangle$. Here, the average extends over molecules of a given topology (LP or CP), and $\tau$ is a dummy time variable. The diffusion constant $D$ was calculated from the slope of the mean-squared displacement according to the relation, $dg_{3}(t)/dt = 6 D$.

\begin{table}
\begin{center}
\begin{tabular}{ccccc}
\hline\hline 
\boldmath{$\phi_{L}$} & \boldmath{$R_{L}$ $^\dag$} & \boldmath{$\sqrt{\frac{g_{3}(\tau_{sim})}{R_{L}}}$} & \boldmath{$R_{C}$} & \boldmath{$\sqrt{\frac{g_{3}(\tau_{sim})}{R_{C}}}$}  \\
\hline
\multicolumn{5}{c}{\textbf{N=150}} \\
\hline
0.500 & 7.95 $\pm$ 0.10 &     & \\
0.479  &  8.21 $\pm$ 0.07  &  3.32  &  5.85 $\pm$ 0.16  &      4.09  \\
0.458  &  8.10 $\pm$ 0.07  &  3.30  &  5.82 $\pm$ 0.11  &      3.63  \\
0.438  &  8.19 $\pm$ 0.08  &  3.23  &  5.76 $\pm$ 0.10  &      3.65  \\
0.375  &  7.99 $\pm$ 0.08  &  3.15  &  5.80 $\pm$ 0.07  &      4.37  \\
0.313  &  8.02 $\pm$ 0.09  &  2.83  &  5.50 $\pm$ 0.05  &      3.97  \\
0.250  &  8.21 $\pm$ 0.10  &  2.39  &  5.54 $\pm$ 0.05  &      3.43  \\
0.188  &  8.16 $\pm$ 0.12  &  2.32  &  5.46 $\pm$ 0.04  &      3.90  \\
0.125  &  8.15 $\pm$ 0.15  &  2.50  &  5.30 $\pm$ 0.03  &      4.59  \\
0.063  &  8.01 $\pm$ 0.18  &  2.58  &  5.25 $\pm$ 0.03  &      5.57  \\
0.042  &  8.04 $\pm$ 0.25  &  2.68  &  5.25 $\pm$ 0.03  &      5.81  \\
0.021  &  8.33 $\pm$ 0.34  &  2.49  &  5.16 $\pm$ 0.03  &      6.32  \\
0.000 &    &  & 5.09 $\pm$ 0.14 & \\
\hline
\multicolumn{5}{c}{\textbf{N=300}} \\
\hline
0.500 & 11.20 $\pm$ 0.17 &     & & \\
0.450  &  11.32 $\pm$ 0.15  &  3.00  &  8.60 $\pm$ 0.20  &     2.64  \\
0.375  &  11.51 $\pm$ 0.18  &  2.89  &  8.71 $\pm$ 0.20  &     2.73  \\
0.250  &  12.27 $\pm$ 0.25  &  2.68  &  8.10 $\pm$ 0.11  &     3.41  \\
0.167  &  12.02 $\pm$ 0.29  &  2.50  &  7.50 $\pm$ 0.11  &     3.77  \\
0.125  &  11.80 $\pm$ 0.31  &  2.40  &  7.32 $\pm$ 0.10  &     4.12  \\
0.100  &  12.37 $\pm$ 0.33  &  2.70  &  7.24 $\pm$ 0.08  &     4.86  \\
0.050  &  12.35 $\pm$ 0.55  &  2.75  &  7.10 $\pm$ 0.07  &     6.27  \\
0.025  &  11.67 $\pm$ 0.56  &  2.79  &  6.96 $\pm$ 0.05  &     7.41  \\
0.000 &      &    & 7.02 $\pm$ 0.06 & \\
\hline

\end{tabular}
\caption{Description of the systems simulated. Simulation box size $L_{box}=60$, and total density $\phi_{C} + \phi_{L} = 0.5$. The radii of gyration $R_{L}$ and $R_{C}$ are reproduced from ref. \cite{iyer07} for reference. The third and fifth columns show that the molecules have diffused at least two times their radius of gyration during the simulation. \label{tab_systems} }
\end{center}
\end{table}

\section{Results}

\begin{figure}
\begin{center}
\includegraphics[scale=0.5]{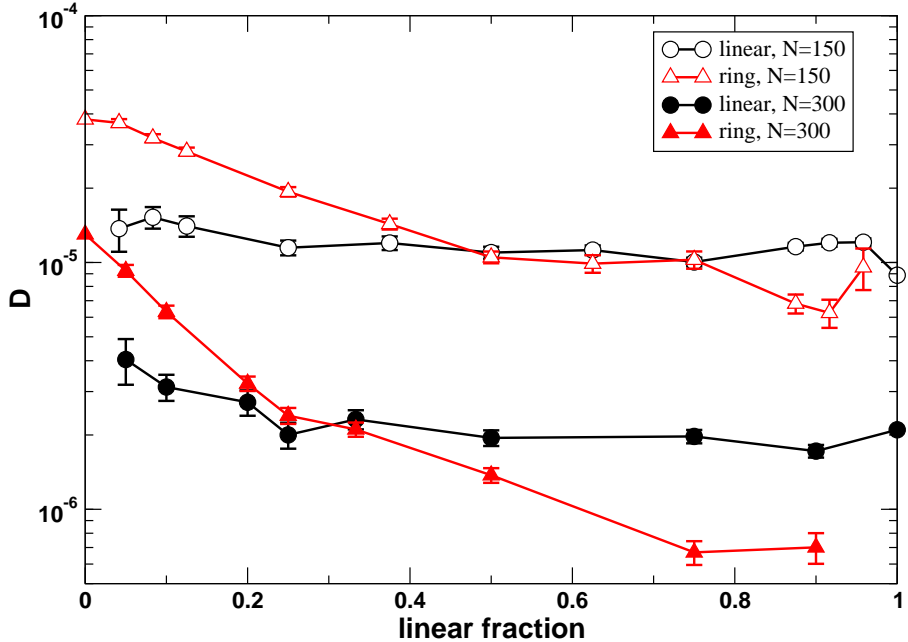}
\caption{The self-diffusivity of the cyclic (triangles) and linear (circles) molecules for $N=150$ (open symbols) and $N=300$ (filled symbols) at different blend compositions. \label{Dsim}}
\end{center}
\end{figure}

To estimate the diffusion constants we performed linear regression analysis on the mean-squared displacement $g_{3}(t)$ in the interval $t=0.15\tau_{sim} - 0.7\tau_{sim}$, where the function $g_{3}(t)$ was almost perfectly linear. Data on pure CPs and LPs was obtained from literature \cite{shaffer94, brown01}. Figure \ref{Dsim} shows the variation of the diffusivity as a function of the fraction of the LPs for $N=150$ and $N=300$, respectively. As the linear fraction increases, the diffusivities of both, the CPs and LPs  decrease, although $D_{C}$ drops more sharply. This is particularly evident for $N=300$. Further, that decrease  is most pronounced at small  $\phi_{L}$.

Experimental data on entangled DNA solutions (fig. 3 from ref. \cite{robertson07b}) suggest that LPs diffuse more sluggishly in a CP matrix than in a LP matrix, and that the strength of this slowdown increases with $N$. This is possibly the reason why the decrease in $D_{L}$ at small $\phi_{L}$ is more prominent for $N=300$. Other studies on polystyrene melts (fig. 2 from ref. \cite{tead92}) suggest that $D_{L}$ is independent of the composition of the blend, which appears to be true over a wide composition range in figure \ref{Dsim}. The apparent contradiction in these two datasets may be partially reconciled through our findings. In all likelihood, a small fraction of the supposed polystyrene CPs in the matrix were contaminated with LPs,\cite{tead92} increasing the actual $\phi_{L}$. It is then conceivable that in the window of observation the decrease in $D_{L}$ at small $\phi_{L}$ was not captured.

\subsection{Minimal Model}
In order to interpret these simulation results, we suggest a minimal model, which ignores prefactors and other numerical details. In this scheme, we visualize a CP in a blend, which is threaded by $Z_{C}$  surrounding LPs. We recall from previous simulations that the average number of entanglements on a CP,  $Z_{C}$, varies according to the linear fraction as $Z_{C}(\phi_{L})/Z_{L} = \phi_{L}/\phi$, where $Z_{L} \approx N/30$ is the average number of entanglements on a LP and is independent of blend composition \cite{subramaniam08}.

The threaded LPs restrain the mobility of the CP. As some of the LPs venture out, others arrive and form entanglements at the same rate, and the equilibrium structure of the melt is not disturbed. Consequently, the primitive path of the CP itself undergoes local rearrangement
as it relaxes by constraint release (CR) Rouse motion \cite{graessley82}. One can argue that,
\begin{equation}
\tau_{C}(\phi_{L}) = \tau_{C}(\phi_{L}=0) + \tau_{CR}(\phi_{L}),
\label{modeleqn0}
\end{equation}

\noindent
where $\tau_{C}(\phi_{L})$ and $\tau_{C}(\phi_{L}=0)$ are characteristic timescales for the motion of a CP in a blend and in a pure melt (no LPs), respectively, and $\tau_{CR}$ is the characteristic CR timescale. When $\phi_{L} \approx 0$, it follows that $\tau_{CR} \approx 0$, and hence $\tau_{C}(\phi_{L}) \approx \tau_{CR}(0)$. Similarly, when $Z_C \gg 1$,  we expect $\tau_{CR}(\phi_{L}) \gg \tau_{C}(\phi_{L}=0)$, and $\tau_{C}(\phi_{L}) \approx \tau_{CR}(\phi_{L})$. The characteristic timescales for CPs and LPs may be approximated by $\tau_{C}(\phi_{L}) \sim R_{C}^{2}(\phi_{L})/D_{C}(\phi_{L})$, and $\tau_{L}(\phi_{L})=R_{L}^{2}(\phi_{L})/D_{L}(\phi_{L})$, respectively. Using standard CR arguments, the local hopping time for the $Z_{C}$ ``effective Rouse" beads (entanglements on the CP primitive path) is set by the LP relaxation time $\tau_{L}$, and one obtains the familiar $\tau_{CR} \sim \tau_{L} Z_{C}^2$. However, this does not complete the description of the CR process, because it assumes that the frictional drag per entanglement segment of the CP, or the effective Rouse bead, $\zeta_{Rouse}$ is a constant. As mentioned earlier, for a CP of a given length $N$, the number of Rouse beads $Z_{C}$ varies with the composition of the CLB. If $\zeta$ is the monomeric friction coefficient, then the total frictional drag of the CP is $\zeta N$, which is distributed among the $Z_{C}$ effective Rouse beads. Therefore, $\zeta_{Rouse}=\zeta N/Z_{C}$, and ignoring constants,
\begin{equation}
\tau_{CR} \sim \tau_{L} Z_{C}^2 \frac{\zeta_{Rouse}}{\zeta N} \sim \tau_{L} Z_{C}
\end{equation}

Thus, equation \ref{modeleqn0} can be rewritten as,
\begin{equation}
\frac{R_{C}^{2}(\phi_{L})}{D_{C}(\phi_{L})} = \frac{R_{C}^{2}(\phi_{L}=0)}{D_{C}(\phi_{L}=0)} + Z_{C} \frac{R_{L}^{2}(\phi_{L})}{D_{L}(\phi_{L})}.
\label{modeleqn}
\end{equation}
Further, $R_{L}(\phi_{L})/R_{C}(\phi_{L})$ is a weak function of both $N$ and $\phi_{L}$, and is easily dominated by the change in self-diffusivity \cite{iyer07, subramaniam08}. Neglecting this change in the size,  we rearrange eqn. \ref{modeleqn} as,

\begin{equation}
Z_{C} \frac{ D_{C}(\phi_{L}=0)}{D_{L}(\phi_{L})} = c_{1} \frac{D_{C}(\phi_{L}=0)}{D_{C}(\phi_{L})} + c_{2}
\label{modelfin}
\end{equation}

\noindent
where $c_1$ and $c_2$ are constants that account for prefactors, ignored in this minimal model.

\begin{figure}
\begin{center} 
\includegraphics[scale=0.5]{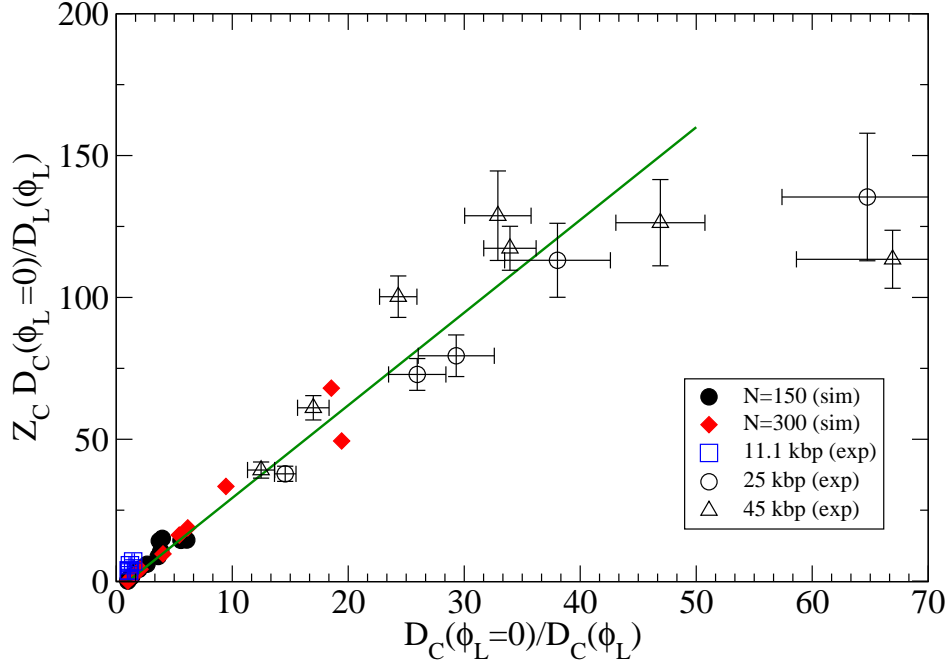}
\caption{To validate equation \ref{modelfin}, the self-diffusivity simulation data (solid symbols) on $N=150$ (circles) and $N=300$ (diamonds) and experimental data (open symbols) on the self-diffusivity of CPs in an entangled linear DNA matrix (ref \cite{robertson07b}) at different concentrations, and three different molecular weights are replotted. The straight line confirms that there is a linear relationship between the quantities on the  horizontal and vertical axes.
\label{sec}}
\end{center}
\end{figure}

From our simulations (fig. \ref{Dsim}), and from prior primitive path analysis,\cite{subramaniam08} all the parameters in eqn. \ref{modelfin} can be determined, and the viability of the minimal model can be ascertained. As shown in figure \ref{sec}, all the available simulation data, independent of composition and molecular weight, collapse on to a linear master curve. 

To test whether experimental data may also abide by eqn. \ref{modelfin}, we employed recently published self-diffusivity data on solutions of linear and cyclic DNA \cite{robertson07b}. In this study, the authors used fluorescence microscopy to  measure the diffusivities $D_{ij}$ of tracer LPs and CPs, in a matrix of either LPs or CPs, where the subscripts $i$ and $j$ represent the topologies of the tracer and matrix molecules, respectively. These diffusivities were compiled as a function of contour length (5.9-kbp, 11.1-kbp, 25-kbp and 45kbp), and total solution concentration (up to 1 mg/ml), and for small values of these two variables the systems were not entangled. Using data on $D_{LL}$ (linear tracer in a linear matrix), we demarcated the transition from unentangled to entangled dynamics. Thus, from figures 2 and 3 in ref. \cite{robertson07b}, we found that the 5.9-kbp sample is too short to be entangled, at any concentration. The concentrations at which the shift to reptation dynamics is observed for the pure LPs of length 11.1-kbp, 25-kbp and 45kbp samples, was 0.7 mg/ml, 0.6 mg/ml and 0.4 mg/ml, respectively. For the most well-entangled systems, they observed $D_{CC} > D_{LC} > D_{LL} > D_{CL}$, which is in accordance with simulation data on $N=300$. Thus, there are 16 data-points available in the entangled regime at different lengths and concentrations for a tracer CP in a LP matrix. Under these conditions, $D_{C}(\phi_{L}) = D_{CL}$, and $\phi_{L}/\phi  \rightarrow 1$. Similarly, $D_{C}(\phi_{L}=0)=D_{CC}$, and $D_{L}(\phi_{L}) = D_{LL}$.

The total concentration $c$, and contour length $l$ contribute differently to the overall dynamics \cite{rendell87}. In the present analysis, the relevant relationships are the dependence of the average number of entanglements per chain on these two parameters. We recall that at a given $c$, that the average number of entanglements per polymer chain is proportional to its contour length, $Z_{L} \sim l$. Similarly, for a given $l$, the entanglement density increases with $c$. Since the modulus $G  \sim c^{7/3}$, and the entanglement length $l_e \sim cRT/G \sim c^{-4/3}$, it follows that $Z_L \sim l/l_e \sim c^{4/3}$ \cite{colby90}. Thus, in the present case, $Z_{C} =Z_{L}= (l/l_0) (c/c_0)^{4/3}$, where $l_0$=3 kbp, and $c_0$=1 mg/ml were chosen,\cite{robertson07b} although it should be pointed out that different values for the entanglement molecular weight of DNA solutions at $c_0=$1 mg/ml between $l_0=$ 1-30 kbp are supported in the literature \cite{mason98, goodman02, boukany08}. If we superpose these data-points (fig. \ref{sec}), we find that although two of the 16 data-points diverge from the trendline, the rest of the data are in very good agreement with the predicted linear dependence.

\section{Summary}

We performed Monte Carlo simulations to estimate the self-diffusivity of entangled CPs and LPs in blends, and constructed a simple constraint release model. Both, the simulation, and experimental data on entangled DNA solutions appear to obey the minimal model over a wide range compositions, concentrations, and molecular weights.

 \end{document}